# Swirling topological textures of polarization in bulk relaxor ferroelectrics

M. Eremenko[1,2,3], V. Krayzman[1,3], S. Gorfman[4], A. Bosak[5], H. Y. Playford[6], P. A. Chater[7], B. Ravel[1], W. J. Laws[1], F. Ye[2], A. Minelli[2], B.-X. Wang[8], Z.-G. Ye[8], M. G. Tucker[2], I. Levin[1*]

[1]Materials Measurement Science Division, National Institute of Standards and Technology Gaithersburg MD 20899, USA

[2]Spallation Neutron Source, Oak Ridge National Laboratory, Oak Ridge, TN, USA

[3]Theiss Research, La Jolla, CA 92037 USA

[4]Department of Materials Science and Engineering, Tel Aviv University, Tel Aviv, Israel

[5]European Synchrotron Radiation Facility, BP 22, 38043, Grenoble, Cedex, France

[6]ISIS Facility, Science and Technology Facilities Council, Didcot, OX11 0QX, Oxfordshire, UK

[7]Diamond Light Source, Science and Technology Facilities Council, Didcot OX11 0DE, Oxfordshire, UK

[8]Department of Chemistry and 4D LABS, Simon Fraser University, Burnaby, British Columbia V5A 1S6, Canada

[*]Email: igor.levin@nist.gov

**Abstract**

A complete understanding of the mechanisms for dielectric relaxation in relaxor ferroelectrics remains elusive. We used a structural refinement framework that integrates several types of experimental data to identify the nanoscale correlations of polarization and their relationship to the underlying chemistry in the classic relaxor system $PbMg_{1/3}Nb_{2/3}O_3$-$PbTiO_3$. The polar structure in these materials in their bulk cubic state can be represented as *overlapping* anisotropic volumes, each encompassing unit cells with projections of their polarization vectors onto the volume's longest axis pointing in the same direction. The overlap results in swirling topological textures of polarization containing vortices, such as merons, and displaying smooth changes in the polarization directions. The locations of these vortices are linked to the electric charge gradient caused by compositional heterogeneities, deemed to create depolarizing fields.

Local perturbations of atomic order caused by competing chemical bonding interactions or chemical heterogeneities are increasingly acknowledged to impact the functional properties of various solid-state materials. Relaxor ferroelectrics, or relaxors, are regarded as the epitome of this phenomenon. Relaxors are distinguished by a broad frequency-dependent maximum of dielectric permittivity, often linked to chemical disorder [1]. These materials enable commercially viable formulations with exceptional electromechanical and energy-storage properties [2-4]. Extensive research has been conducted to relate such properties to the underlying chemistry and structure, producing a general theoretical framework and empirical relationships currently guiding the materials' development [5]. The relaxor behavior is widely accepted to be caused by local variations in chemistry which create polar inhomogeneities perceived as defining the dielectric response. However, a seven-decade quest to understand or at least adequately describe these inhomogeneities has been unsuccessful, with this task remaining a fundamental challenge [6].

A canonical example of such systems is the perovskite-like relaxor PbMb$_{1/3}$Nb$_{2/3}$O$_3$ (PMN) and its solid solutions with PbTiO$_3$ (PT) [2], which demonstrate outstanding electromechanical performance. This performance of PMN-PT, especially in the single-crystal form, is attractive for medical imaging, sonar, and other sensor and actuator applications. Numerous attempts to establish the relationship between the structure and properties in PMN-PT suggested several, sometimes seemingly contradictory, scenarios but no ultimate clarity [6]. The proposed depictions of the polarization vary from isolated polar nanoregions (PNRs) embedded in a non-polar matrix [7] to polar nanodomains separated by domain walls [8] to PNRs separated by wide boundaries featuring gradual changes in polarization [9-11]. Still, some works questioned even the very concept of PNRs and their significance in explaining the relaxor properties [12]. A principal reason for this ambiguity is the inherent multilevel complexity of the relaxor structures, which combine chemical disorder with dynamic and static polar displacements that evolve upon changing temperature [13-14]. Yet, each measurement method can capture only certain aspects of this complexity, making it difficult to reconstruct a three-dimensional (3D) polarization vector field and its relationship to the underlying chemistry over all the relevant length and time scales.

It is well established that PMN maintains an average cubic structure even at very low temperatures. In this state, Mg and Nb octahedrally coordinated by oxygen undergo partial rock-salt-type ordering on the nanoscale. Recent results [10, 15] have confirmed earlier inferences from dark-field transmission electron microscopy (TEM) [16] of the nanoscale regions with stronger ordering separated by those featuring a gradient of the order parameter. The partial disorder of Mg and Nb frustrates the off-center shifts of Pb atoms, which are stabilized by the hybridization of the Pb 6$s^2$ and O 2$p^6$ electrons and contribute significantly to the polarization. Refinements of the PMN structure using large-scale atomic configurations revealed a hierarchical assembly of PNRs featuring cooperative cation displacements relative to the oxygen framework [10]. The neighboring small-scale nanoregions with the polarization aligned parallel to the ⟨111⟩ directions were predominantly ≈71° variants. The agglomerates of several of these nanoregions exhibiting the head-to-tail arrangement of their polarization vectors appeared to form larger-scale PNRs.

The essential characteristics of structural snapshots from these refinements [10] agreed with molecular dynamics (MD) simulations [9]. Both sets of results indicated a lack of a non-polar matrix, instead revealing PNRs separated by regions with a gradually changing direction and magnitude of polarization akin to wide, low-angle domain walls. Atomic-resolution scanning TEM images (STEM) [11] supported a multi-domain state with a high density of such domain walls, with their locations pinned by structural heterogeneities, such as regions of stronger chemical ordering or areas with larger octahedral distortions and tilting. The large-box structural refinements [10] and MD simulations [9] agree on a multi-domain state responsible for the observed anisotropic X-ray and neutron diffuse scattering (DS). Still, the specific correlation topology contributing to this anisotropy remained unclear. The proposed hierarchical assembly of PNRs separated by wide, low-angle boundaries hints at the existence of inter-PNR correlations. However, such correlations have not been considered, even though they could be essential for clarifying the mechanisms of excellent electromechanical properties in PMN-PT crystals.

Here, we addressed this question by explicitly recovering interatomic correlations over the ten-nanometer length scale in PMN and PMN-PT compositions (with 30 % and 35 % PT) in their high-temperature cubic state. Our approach involved refining large-scale structural models simultaneously against multiple types of experimental data, including neutron and X-ray total scattering, extended X-ray absorption fine structure, and 3D reconstructions of X-ray DS. Advancements in data-analysis techniques and refinement software achieved as a part of this work permitted a significantly deeper understanding of the relaxor structure than was previously achievable.

Our findings provide evidence for three primary effects not covered by the existing models. First, the nanoscale chemical ordering is enhanced in regions with a higher content of larger, weakly polarizable Mg ions. These regions disrupt longer-range correlations among polar displacements and alter the polarization vector field. Second, cation displacement correlations are multi-level, with correlation volumes that extend across multiple PNRs. Third, the inter-PNR correlations yield 3D labyrinth-like textures of *continuously* curling polarization that contain vortex-like formations, such as merons, acquiring net dipole moments. The development of these topological textures is attributed to the mitigation of the depolarizing electric fields that arise from the nanoscale compositional and charge heterogeneities.

*Local ordering vs compositional fluctuations*

We focused on the cubic ergodic state below the so-called Burns temperature [17], which signifies the onset of extended correlations among polar displacements. This choice allows for a straightforward comparison of different PMN-PT compositions, emphasizing the crystal-chemical trends that are precursors to the lower-temperature (non-ergodic) states. Additionally, it helps to avoid the refinement complexities in cases of monoclinic and tetragonal twinning encountered in PMN-PT at lower temperatures [18]; an effective methodology for such refinements is still under development. Details of refinements and the fitting results are discussed in the Methods section and SI (figs. S1-S13).

In PMN, the short-range rocksalt-type cation order (Fig. 1a, fig. S14) is similar to previous reports [10]. The refined order parameter and correlation length differ slightly from the published values due to our more accurate treatment of the diffuse intensity scale and width. For PMN-PT compositions with three B-site species, a description of ordering requires three independent short-range order parameters [19]. The results are summarized in Tables S2 and S3 and fig. S15. The rocksalt type ordering for Nb and Mg is retained, but its degree and spatial extent are significantly reduced compared to PMN.

In PMN, the local order parameter (see Methods) for Mg and Nb is closely linked to the local Mg content, peaking for a 1:1 ratio of the two B-site species (Fig. 1b). That is, the regions with stronger ordering are enriched with Mg, leading to the enhanced and more extended nanoscale fluctuations in lattice strain and electric charge (Fig. 1c) compared to a random distribution of the B-site cations [20].

As demonstrated in prior works [10, 14], the magnitude of Pb displacements in PMN increases linearly as the number of Mg atoms, $n$, in the [PbMg$_n$Nb$_{8-n}$] clusters increases; this trend is driven by the bonding requirements of the oxygen atoms. The same behavior occurs in the PMN-PT compositions. For a given local Mg/Nb ratio, the displacements are more significant for a larger number of Ti around Pb. While both the local order parameter and the magnitude of the Pb displacements scale with the local Mg concentration, the direct statistical correlation between the two is weak.

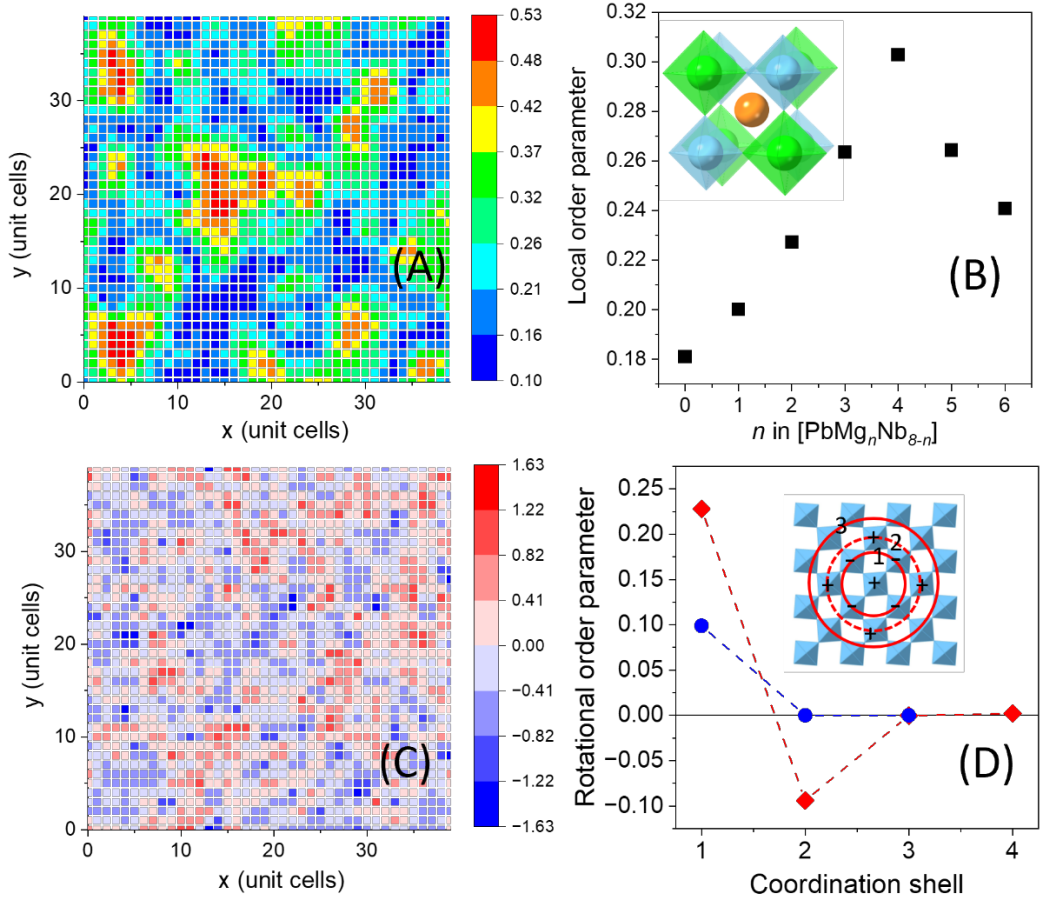

Fig. 1: **Nanoscale chemical ordering and compositional segregation**.
(A) Single layer of Pb in PMN with the color scale reflecting the average local B-site (Mg, Nb) order parameter in the [PbB$_8$] coordination clusters centered on the A-site Pb cations in this layer. This local order parameter was calculated for each B-cation (see Methods) and then averaged over the eight B cations surrounding each Pb ion. (B) Relationship between the local B-cation order parameter around Pb atoms and the number of Mg cations in the [PbMg$_n$Nb$_{8-n}$] coordination clusters. Inset shows an example cluster with $n=4$ and complete rocksalt ordering of Mg (green) and Nb (blue). (C) Same layer of Pb as in (B) but with the color scale reflecting the net formal electric charge (in $e$) for each Pb-centered unit cell. This charge varies with the local Mg$^{2+}$/Nb$^{5+}$ ratio and the map reveals the extended, positively (Nb$^{5+}$-rich) and negatively (Mg$^{2+}$-rich) charged regions. (D) Local order parameter for octahedral rotations in PMN within layers of [BO$_6$] octahedra around the axis normal to the layer plane. (red) 300 K; (blue) 490 K. The inset illustrates a perfectly ordered pattern of rotations maintaining the octahedral rigidity.  Signs of rotations are indicated using "+" and "-" symbols.  Circles label successive coordination shells for the octahedron in the center.

The oxygen framework accommodated the differences in the ionic radii of $Mg^{2+}$ (0.72 Å), $Nb^{5+}$ (0.64 Å), and $Ti^{4+}$ (0.605 Å) [21], with the [$MgO_6$] octahedra expectedly expanding and the [$NbO_6$] (weak change) and [$TiO_6$] octahedra contracting relative to the average (fig. S16).  The magnitude of this deformation scales with the local chemical order parameter: stronger ordering allows for larger oxygen shifts along the B-O-B' bonds. The PMN structure at 300 K contained small clusters (1 nm to 2 nm) with the in-phase $a^0a^0c^+$-type [21] octahedral tilting (Fig. 1D), with the rotation angle ≈4°.  At 490 K, the degree of rotational ordering was reduced, and the coherency length of these tilts was limited to the nearest-neighbor octahedra.  No rotational ordering was observed in PMN-PT compositions.

*Correlations among polar displacements:  correlation length vs. PNR size*

In PMN at 300 K and 490 K and in PMN-PT at 533 K, the Pb ions are offset from their average cubic position (figs. S17- S19).  However, for the PMN-PT configurations, the magnitude of the Pb off-center shifts is reduced compared to PMN (fig. S19), in line with the shrinkage of the average lattice (fig. S1). The preferred directions of Pb displacements also change from ⟨111⟩ and ⟨100⟩ for PMN to ⟨100⟩  for PMN-PT. (fig. S20).  Regardless of composition, the displacements of Nb favor the ⟨111⟩ (primary) and ⟨110⟩ directions.  In contrast,  the Ti cations in PMN-PT are preferentially shifted along ⟨100⟩ (fig. S21 and fig.S22).

The local alignment of cation displacements can be characterized using the average angle, $\alpha$,  between the displacement vector of a given cation and those of its nearest cation neighbors [10] (e.g., coordination environments [$PbPb_6$], [$PbB_8$] or [$BPb_8$]).  The angles $\alpha$≈0° and $\alpha$≈180° indicate parallel and anti-parallel alignments.  An atomic configuration with randomly oriented displacements yields ⟨$\alpha$⟩≈90° (here, the angular brackets denote the configuration average).

For [$PbPb_8$] in PMN, we obtained ⟨$\alpha$⟩≈77° and ⟨$\alpha$⟩≈82° at 300 K and 490 K, respectively, indicating a preference for the parallel alignment (from our experience and testing, any deviations from ⟨$\alpha$⟩=90° greater than  5° are deemed significant).  Likewise, for the PMN-PT configurations, ⟨$\alpha$⟩≈80.°  Dense clusters of atoms with small values of $\alpha$ (e.g., $\alpha$<45°) can be regarded as PNRs [10]. In PMN, at 300 K, for the [$MgPb_8$] clusters, ⟨$\alpha$⟩≈88°, whereas for [$NbPb_8$], ⟨$\alpha$⟩≈73° – consistent with a weakly polarizable nature of $Mg^{2+}$ ions which are much less effective in supporting displacement correlations than smaller, higher-charge species, like $Nb^{5+}$.   For the PMN and PMN-PT configurations, the $\alpha$-values for the [$PbPb_6$] clusters scale positively with those for the [$PbB_8$] units centered on the same Pb atoms (fig. S23), suggesting that better alignment of displacements between Pb and its neighboring B-cations (distance=$a\sqrt{3}/2$, where $a$ is the lattice parameter) enhances the alignment within the Pb-Pb coordination sphere (distance=$a$).

We are interested in spatial correlations among atomic displacements, which, as refined, are obscured by randomness associated with thermal disorder and the nature of the refinement procedure.  We extracted correlated components of displacements using the Fourier filtering of the *calculated* DS *amplitude*, with regions of interest selected by placing spherical masks around all Bragg reflections (see Methods). This filtering highlights regions featuring the patterns of atomic displacements or chemical ordering giving rise to specific diffuse features in diffraction space. Indeed, after applying the filter, the correlations are seen much more clearly (Fig. 2A, B). In PMN at 300 K, the filtered components for Pb yield ⟨α⟩≈21°, indicative of a much stronger local alignment compared to the total displacements with ⟨α⟩≈77°. Those Pb atoms that contribute the most to the DS, as reflected in their acquisition of the largest intensity in the inverse Fourier transform of the DS amplitude, have Mg-deficient local environments with a lesser degree of the B-cation ordering. That is, higher local concentrations of Mg weaken the correlations among the Pb displacements, consistent with Mg acting as a "blocking" species [23].

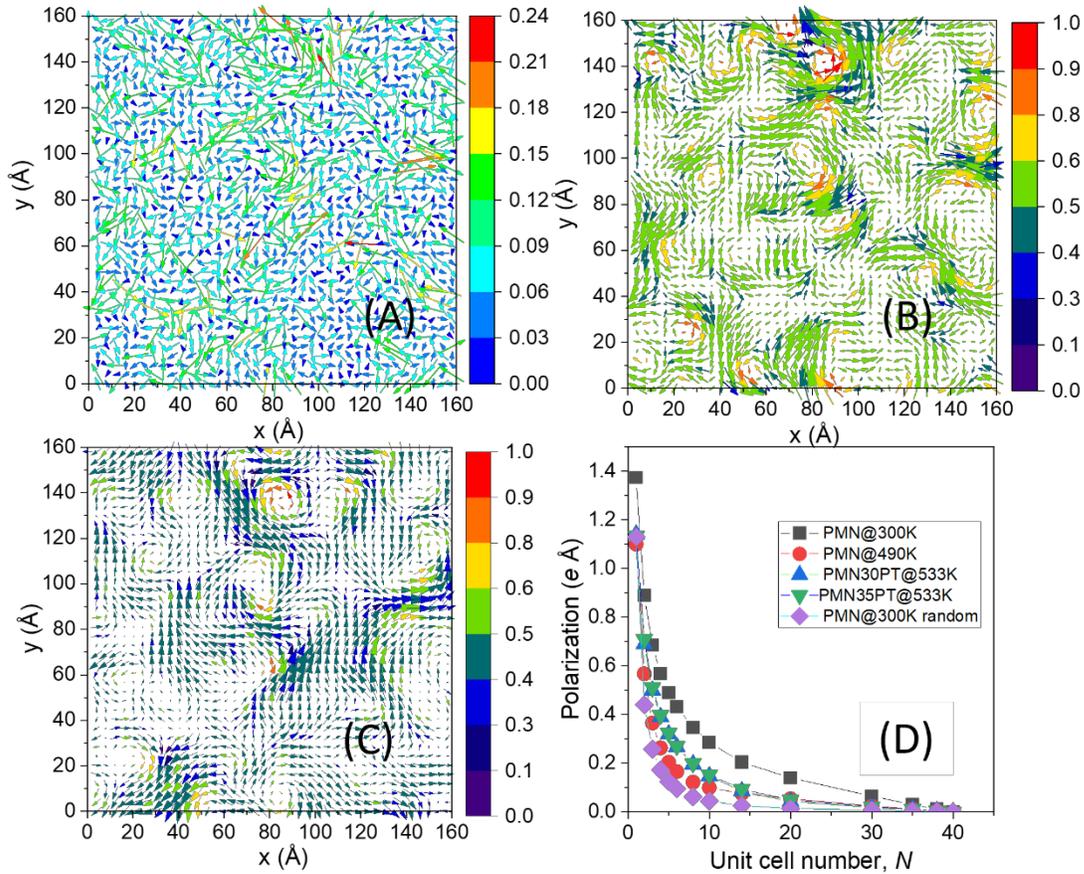

Fig. 2: **Projected vector fields revealing polar correlations**.
(A) Projection of the Pb columns in the refined PMN (300 K) configuration onto the (001) plane with vectors indicating column displacements. Color – displacement magnitude in Å. (B) Same as (A) but for a configuration with Pb displacements derived via the Fourier filtering of the calculated diffuse scattering amplitude. Color – the $Q$-criterion describing the vorticity of the 2D vector field (see Methods). (C) Projection map similar to (A) and (B) but with the vector field corresponding to the local polarization calculated from the refined coordinates of all the atoms in the configuration and formal ion charges (see Methods). Color – the Q-metric as in (B). The patterns in (B) and (C) differ somewhat but are strongly correlated. (D) Dependence of the local polarization on the averaging volume of $N \times N \times N$ unit cells for PMN at 300 K and 490 K and for 0.7PMN-0.3PT and 0.65PMN-0.35PT at 533 K. The dependence for a spatially random distribution of the local polarization in PMN at 300 K is shown as a reference (here, the magnitudes and directions of the polarization vectors are the same as in the refined configuration, but any spatial correlations are scrambled). The polarization map shown in (C) corresponds to *N*=2.

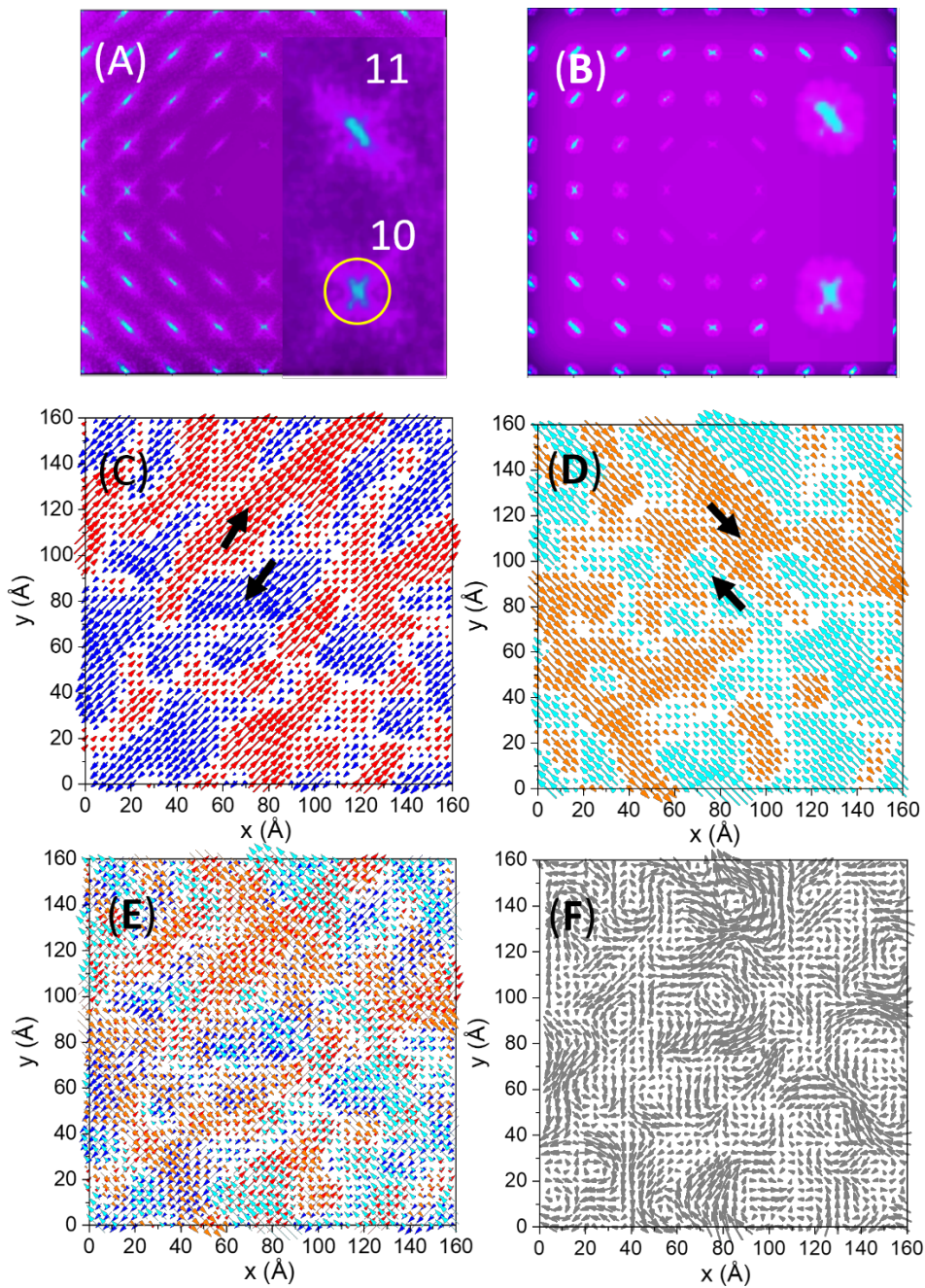

**Fig. 3: Inter-PNR correlations and origins of diffuse scattering**.
(A) and (B) Diffuse intensity calculated for the projection maps in Fig. 2A and Fig. 2B, respectively. In both panels, insets show a magnified view of the diffuse intensity around the [10] and [11] Bragg peaks (indexes are in 2D). The circle in the inset in (A) illustrates the size of the mask used in the Fourier filtering. (C) and (D) [11] and [1-1] components, respectively, of the displacements in Fig. 2B and Fig.3F. In each panel, the two colors highlight the positive and negative displacement directions. (E) Overlay of

(C) and (D). (F) Same map as in Fig. 2B to facilitate the comparison with (E). The vector sum of the two components in (E) can be seen reproducing the total displacement in (F).

We found that the array of the correlated Pb displacements can be represented as a 3D assemblage of overlapping regions, which, if approximated as ellipsoids, exhibit one principal axis significantly longer than the other two. The Pb displacements within each ellipsoid are parallel to its long axis. The overlap of such volumes creates a 3D maze of smaller PNRs. The displacements within the PNRs that belong to several volumes represent a vector sum of the corresponding components.

This picture can be illustrated using two-dimensional (2D) displacement fields for the projections of Pb columns onto the {001} planes as shown in Fig. 3 for PMN at 300 K; similar fields albeit of purely static displacements would be revealed in atomic-resolution (S)TEM images. Such projections (Fig. 2A, B) generate 2D DS (Fig. 3A and B) with an anisotropy similar to that observed in 3D. The ±[11] (2D indices) and ±[1-1] displacement components (Fig. 3C, D) of the column displacements (Fig. 3E) each display correlations that yield labyrinth patterns with larger correlation lengths and faceting along the [11] and [1-1] directions, respectively. Each component (Fig. 3C and D) features stripe-like clustering of positive and negative directions. Overlapping these patterns results in patch-like domains (Fig. 3F), seen in the total projection (Fig. 3E), with displacements in the overlapping regions being a vector sum of the components. That is, the displacements in the adjacent PNRs are correlated. Simulations based on these patterns show that obtaining a streak-like diffuse intensity around the *HK (H, K $\neq$ 0)* spots requires displacements aligned with the long dimension of the correlated regions; otherwise, this intensity would appear cross-like as seen around the *H*0 and *K*0 peaks. This principle also holds in 3D, where displacement components maintain their correlation across multiple PNRs, with the most extended correlation length along the direction of these displacements.

The ***q***-dependence (vector **q** defines the deviation from a Bragg position) of the DS around the Bragg peaks in PMN at 490 K differs from that at 300 K; however, the tails in the directions transverse to ***Q***=⟨110⟩* remain similarly extended. The substitution of PT into PMN results in more isotropic and narrower distributions of the diffuse intensity, indicating longer correlation lengths than those in pure PMN, in agreement with previous reports [24]. While our refined models capture these changes, the increase in correlation length with higher temperatures or with the substitution of PT appears to contradict the trend for the local polarization (Fig. 2D) calculated directly from the atomic coordinates (see Methods), which exhibits shorter correlation lengths compared to PMN at 300 K. This apparent discrepancy arises because the correlations that define the distribution of the diffuse intensity span several PNRs (Fig. 3). The greater extent of these correlations in PMN-PT can be attributed to a smaller concentration of the blocking Mg species and weaker cation ordering, leading to a weaker Mg segregation. The correlation length of polarization in PMN decreases markedly from 300 K to 490 K. The correlation length in both PMN-PT compositions at 533 K is greater than that in PMN at 490 K; however, it is still substantially smaller than this length in PMN at 300 K.

Previous studies invoked ellipsoidal PNRs in a matrix to explain the anisotropy of the diffuse intensity, with both oblate and prolate shapes proposed [25-26]. However, these models fail to predict the correct **q**-dependence of the diffuse intensity [27]. Our findings reveal a principally different perspective, where such volumes with correlated displacements exist only when considering specific displacement components rather than the total displacements. The shapes of these volumes, which overlap, yielding intertwined PNRs, explain the overall anisotropy of the diffuse scattering but not the complete intensity distribution that includes contributions from inter-PNR correlations.  The size of these "elementary" PNRs, also referred to as $\alpha$-PNRs in [10], which are characterized by having a distinct direction of the total polarization, can be smaller than the correlation length for specific displacement components.  The low-angle $\alpha$-PNR boundaries imaged in TEM [11, 18] are a product of such inter-PNR correlations.

The divergence calculated for a vector field representing the correlated Pb displacements (See Methods) varied with the local B-site chemistry, from preferentially positive for the Nb-rich environments to negative for the Mg-rich ones (fig. S25).  A similar trend was obtained for the local polarization.  This indicates that Nb-rich and Mg-rich regions act as sources and sinks for polarization, respectively, consistent with other experimental findings and simulations [28].  Spatial distributions of the divergence metric for PMN at 300 K and 490 K were correlated, suggesting this effect to be at least partly static, also in line with it being captured by STEM images [28].  In the PMN-PT compositions, both Mg- and Ti-rich clusters exhibited a preferentially negative divergence of the Pb displacements.

The displacements of Nb are aligned with those of Pb (fig. S26), resulting in a relatively narrow Nb-Pb distance distribution.  In contrast, the displacements of Pb and Mg are only weakly correlated, which, combined with the enhanced magnitudes of Pb displacements around Mg, yields a broad and distorted distribution for the Mg-Pb distances.  Interestingly, Ti displacements are *anti-correlated* with those of the neighboring Pb atoms (fig. S26), as in the soft mode in $PbTiO_3$ [29].  This anti-correlation causes Ti to be off-centered within the $[TiPb_8]$ clusters, resulting in a broad and distorted distribution of Ti-Pb distances.

*Inter-PNR correlations and swirling textures of polarization*

A striking feature of the projected filtered Pb displacements is that they form well-defined vortices, with gradual changes in vector orientation near the vortex cores (Fig. 2B). Similar patterns are observed in the local polarization vectors derived directly from the refined atomic coordinates (Fig. 2C).  We calculated various metrics of vorticity (i.e., curl, *Q*-criterion, *etc.*, see Methods) for the 3D displacement field to identify the vortex-like features.  Fig. 4 illustrates a cluster of Pb displacements that spiral around the $\langle 011 \rangle$ direction, resembling a meron tube with a net displacement along its axis [30].  We also observed similar winding textures for the B-cation displacements (fig. S27), consistent with the strong correlations between the two sublattices.

Meron-like formations in PMN (Fig. 4) can be viewed as consisting of correlated PNRs, predominantly pairs with the $\langle 111 \rangle$ and $\langle 001 \rangle$ displacements (angle $\approx 54°$) or 71° $\langle 111 \rangle$ variants (fig S28).  A previous study [10] identified a hierarchical structure, highlighting a fragment where several 71° PNRs create a curling polarization but did not elaborate on the nature of the observed hierarchy.   Our results demonstrate how such "elementary" PNRs, referred to as $\alpha$-PNRs in [10], organize into topological textures.  Streamlines of the displacement vector field reveal a complex correlation topology characterized by continuously curling polarization and multiple vortices identifiable by larger curl-metric values (Fig. 2B, Fig. 4).   Using this metric and HDBSCAN [31], we isolated atoms that form vortex cores, which were predominantly found near the boundaries separating the nanoscale regions with the net positive and negative charges (Fig. 1C) arising from the B-cation distribution. This relationship is visible in 2D slices of the refined configuration (Fig. 5A, B) and is supported by statistical analysis of the distances between the core centers and charge boundaries in three dimensions (Fig. 5C, D). Thus, the vorticity of the displacement field peaks at the maxima of the electric charge gradient, which is defined by the local chemistry.

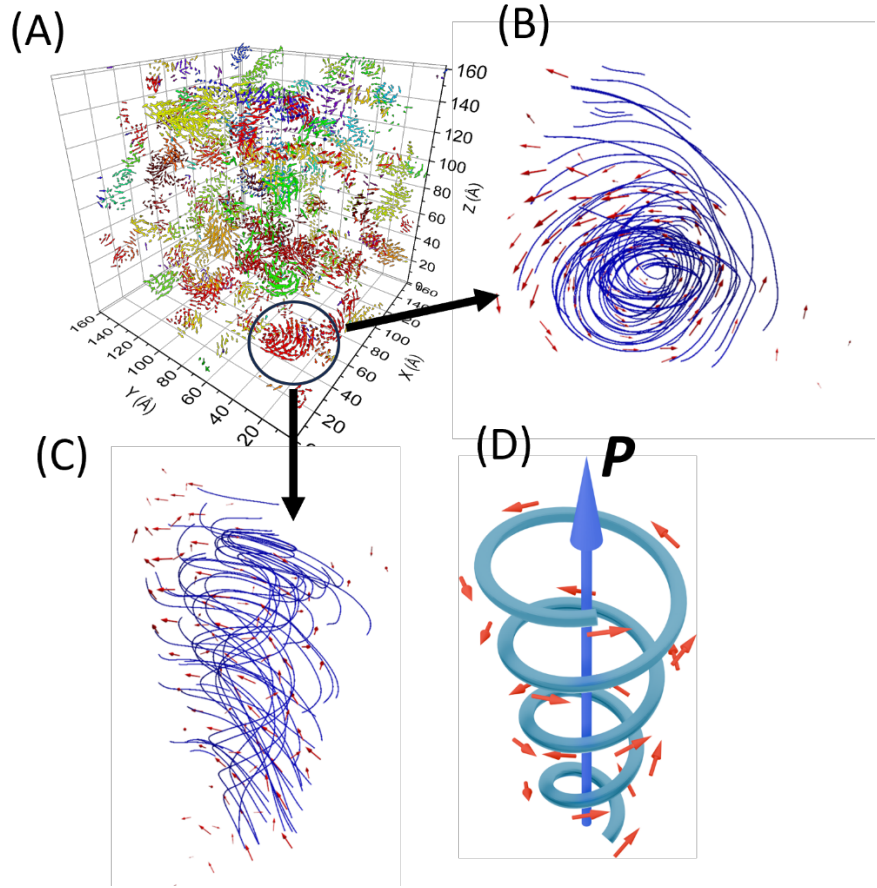

**Fig. 4: Swirling textures of polarization and merons**.
(A) 3D rendering of Pb displacement (filtered) vector field displaying the largest vortex-like clusters. (B), (C) different views of the Pb displacement pattern in one of the well-defined meron-like configurations in (A). Lines are streamlines introduced as a guide to the eye. The tube axis here is approximately parallel to the [011] direction. (D) Schematic rendering of a meron tube with the blue arrow indicating the direction of the net polarization, *P*.

The DS calculated for the Pb atoms with the largest curl metric appeared as isotropic blobs around the Bragg peaks, modifying the intensity dependence on **q**. We could not distinguish between the scattering from the spiraling pattern and other types of correlations. Nonetheless, this result highlights the importance of accurately reproducing the diffuse intensity behavior to capture all the correlations in the displacement field.

The experimental data used in this study and our refined models represent structural snapshots (see methods). Energy-resolved neutron scattering measurements in PMN [32] show that at $E$=0 loss (≈0.5 µeV energy resolution), above 420 K, the elastic DS mostly disappears. Hence, at 490 K, the DS appears to be mainly dynamic in origin, while at 300 K, the static contributions to the diffuse intensity become significant, albeit still non-dominant. The characteristic decay times associated with these dynamics were determined to be of the order of $10^{-1}$ ns to $10^{-2}$ ns [32]. We thus conclude the swirling topology of polarization identified here to be essentially dynamic albeit influenced by chemistry, as the locations of the vortex cores are tied to the B-cation distribution. The observed overlapping correlation volumes for the components of polar displacements are consistent with the phenomenological model of a slowly changing displacement pattern proposed in [27]. It would be interesting to compare the diffuse scattering shown for structural projections in Fig. 2 with that calculated for analogous projections obtained using atomic-resolution STEM images [11, 18, 28]. While projections of our configurations represent structural snapshots, those from STEM highlight purely static displacements.

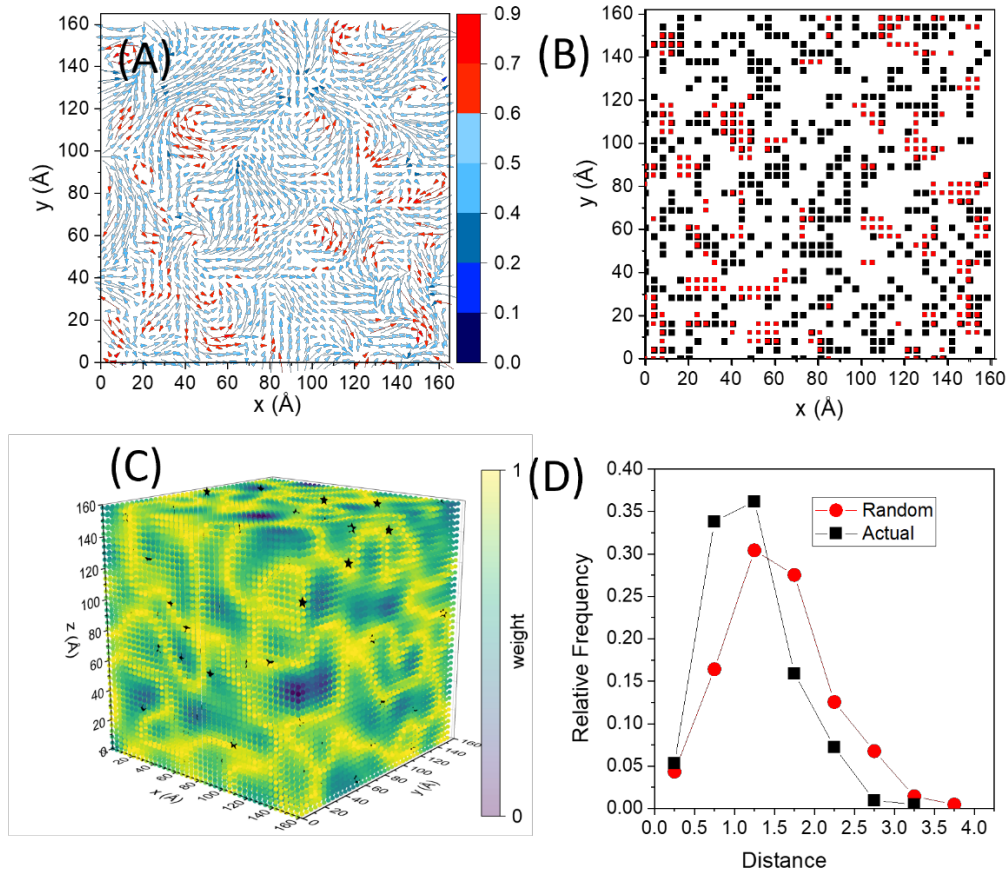

**Fig. 5: Pinning of polarization vortices by the local chemistry.**
(A) Single atomic layer of Pb in PMN, with vectors indicating projections of the filtered Pb displacements onto the layer's plane. Color – Q-criterion of vorticity. The color scale is selected to emphasize the displacements with Q>0.6. (B) An overlay of the Pb atoms having Q>0.6 (red symbols) and locations of the electric charge boundaries (black symbols) in this layer arising from the distribution of the neighboring Mg and Nb. This image reveals a clear preference for the Pb atoms acquiring large Q values to be located in the vicinity of these boundaries. (C) 3D rendering of the charge boundaries (highlighted in yellow) in the refined configuration of PMN with the asterisks marking the centers of Pb clusters having Q>0.6 (identified using HDBSCAN). (D) Statistical distributions of distances from the centers of clusters with large vorticity to the charge boundaries. Black – cluster centers as found in the refined configuration. Red – randomized locations of the cluster centers. The Matt-Whitney test confirms that the actual distances are significantly shorter than those for randomly distributed cluster centers, in line with the visual inference from Fig. 5B.

The topological ordering of polarization in ferroelectrics is of significant interest both fundamentally and technologically [34]. Vortices and bubble domains have been observed in thin-film ferroelectric heterostructures [34] and nanoparticles [35] and interpreted as a mechanism to minimize the otherwise large depolarization field. Levanyuk and Blinc conjectured that bulk relaxors could also undergo a local phase transition to a vortex state [36]. Recent atomic-resolution STEM imaging has provided evidence of bubble domains with skyrmion topologies in bulk NBT-based ceramics [37]. Additionally, simulations for bulk PZT [38] indicate that the topological ordering of PNRs can reduce depolarizing fields. Our finding of swirling polarization textures containing a high number density of vortices in bulk PMN-PT not only reinforces the theoretical predictions but demonstrates the very existence of this phenomenon. The localization of vortices at the charge boundaries created by the nanoscale chemical inhomogeneities observed here suggests that these boundaries act similarly to interfaces or surfaces in synthetic ferroelectric nanostructures, creating depolarizing fields that drive the development of topological textures and defects.

**Summary**

We used an integrative structural refinement framework and a novel software tool to identify the elusive interatomic correlations in PbMg$_{1/3}$Nb$_{2/3}$O$_3$-PbTiO$_3$ (PMN-PT) relaxor ferroelectrics. The previously proposed hierarchical assembly of polar nanoregions (PNR), dominated by the local Pb displacements, has been revealed as overlapping volumes with anisotropic shapes having one principal axis longer than the others. Each volume encompasses atoms with projections of their displacement vectors onto this volume's long axis pointing in the same direction – a correlation that determines the characteristic anisotropy of the observed DS near the Bragg peaks, a hallmark of the PMN-PT system. The overlap of these volumes creates a 3D patchwork of PNRs, with the displacements in the overlapping spaces being a vector sum of the components. The polarization follows these displacement patterns. The nanoscale chemical ordering of the octahedral cations is accompanied by the segregation of Mg into regions with stronger ordering, which enhances the nanoscale compositional heterogeneities. Our results indicate that the compositional effect on polar displacements dominates over that of the ordering. Dense PNRs that contribute the most to the observed DS occur preferentially in Mg-deficient regions, signifying the disruptive role of Mg in establishing polar correlations. The Pb displacements and polarization vector fields are also tied to the chemistry via a divergence that changes from positive to negative between the Nb-rich and Mg-rich regions.

While we invoked PNRs to describe the observed swirling textures of polarization, it was done as a matter of convenience rather than a necessity, similar to a way of rationalizing some topological textures using flux closure domains. However, applying the PNR concept obscures the continuous nature of the changes in the direction and magnitude of polarization, where using vector-field descriptors (e.g., divergences, curl, $\alpha$-metric) could have been more appropriate. That is, the answer to the question posed by Hlinka [12]: "Do we need the ether of polar nanoregions?" appears to be "No, we do not." Takenaka *et al*. [9] drew an analogy between the thermal evolution of relaxors and water, which turns into a slush upon cooling, before freezing. The PNRs were suggested to act like bits of ice in this slush, able to rotate independently from one another. However, we observe a distinctly different scenario where the polar structure is a continuous interrelated maze.

The picture uncovered here shows the polarization vectors aligned over the nanoscale, oscillating around several preferred crystallographic directions. The latter are maintained over a larger length scale before diverging to symmetry-equivalent or other favorable directions that enable smooth changes in the polarization direction, creating swirling textures. We hypothesize that this swirling topology is such that it simultaneously minimizes strain and depolarizing fields. The oscillations and the branching of the nanoscale polarization directions are presumed to arise from the nanoscale chemical heterogeneities discussed above. Indeed, the vortices' cores, characterized by the largest vector curl magnitude, are pinned by the cation chemistry, being located preferentially near the boundaries separating the negatively ($Mg^{2+}$-rich) and positively ($Nb^{5+}$-rich) charged regions. The relaxation times and their statistical distributions for the topological textures of dipole arrays observed here placed under an *ac* electric field are expected to differ from those for uncorrelated PNRs (e.g., cases with a high ratio of blocking over polarizable species). This can be a reason for the differences in the frequency dependency of the dielectric response between PMN and other relaxor systems, e.g., $Ba(Zr,Ti)O_3$-based, which incorporate mixtures of Isovalent species that do not generate depolarizing fields.

A transition from relaxor to classic ferroelectric behavior in PMN-PT aligns with a basic crystal-chemical framework [23], in which the type and concentration of blocking ions influence the correlation length of polar distortions. As the chemical ordering becomes significantly weaker and more limited in scale, the accompanying segregation of Mg is also diminished. The combination of aliovalent $Mg^{2+}$, $Ti^{4+}$, and $Nb^{5+}$ still results in extended charge fluctuations; however, these fluctuations are considerably smaller in magnitude compared to those in PMN (fig. S29. This reduction lowers the barriers to establishing long-range ferroelectric ordering upon cooling. Our analysis indicates that a largely disordered mixture of Ti, Nb, and Mg still affects the directions of lead (Pb) displacements. For instance, Ti behaves similarly to Mg by promoting a negative divergence in the Pb vector field, in contrast to the positive divergence associated with Nb. This influence may explain why the correlated polar disorder observed in the cubic ergodic state persists in the lower-symmetry monoclinic and tetragonal structures of PMN-PT.

**Acknowledgments:** Experiments at the ISIS Pulsed Neutron and Muon Source were supported by beamtime allocation from the UK Science and Technology Facility Council. We thank Diamond Light Source for access to beamline I15-1 and the European Synchrotron Radiation Facility for the provision of beam time on ID28. Portions of this research were carried out at the National Synchrotron Light Source II (NIST beamline 6-BM) and at the Spallation Neutron Source, operated for the DOE Office of Science by Brookhaven National Laboratory and the Oak Ridge National Laboratory under Contracts No. DE-AC02-98CG10886 and DE-SC0012704, respectively. S. G. and I. L. acknowledge the US-Israel Binational Science Foundation for their financial support (Award No. 2018161). S.G. was also supported by the Israel Science Foundation (Awards 1561/18, 1365/23). B.-X.W. and Z.-G.Y. acknowledge the supports for the Natural Sciences & Engineering Research Council of Canada (DG, RGPIN-2023-04416) and the U. S. Office of Naval Research (N00014-21-1-2085).

**Author Contributions:** Conceptualization: I.L , S.G., and M. G.T.  Synthesis:  W. J. L., B.-X.W. and Z-G.Y. Data acquisition and processing: I. L., S.G., H.Y.P., A. B.,  M.G.T., B. R., P. A. C.., F. Y., and A.M. Methodology: M. E., V. K., IL. Software development: M. E and V. K. Data analysis: M. E., V. K., and I.L. Supervision: M.G.T and I.L. Writing – original draft: I.L. and M.E. Writing – review & editing: All coauthors.

**References and Notes**

1. L. E. Cross, Relaxor Ferroelectrics. Ferroelectrics, 76 [3,4] 241-267 (1987)
2. S. J. Zhang, F. Li, High performance ferroelectric relaxor-PbTiO3 single crystals: status and perspective. *J. Appl. Phys.*, 111 (3) 031301 (2012)
3. F. Li, D. B. Lin, Z. B. Chen, Z. X. Cheng, J. L. Wang, C. C. Li, Z. Xu, Q. W. Huang, X. Z. Liao, L. Q. Chen, T. R. Shrout, S. J. Zhang, Ultrahigh piezoelectricity in ferroelectric ceramics by design. *Nature Mater.*, 17 [4], 349 (2018)
4. G. Wang, Z. L. Lu, Y. Li, L. H. Li, H. F. Ji, A. Feteira, D. Zhou, D. W. Wang, S. J. Zhang, I. M. Reaney. Electroceramics for high-energy density capacitors: current status and future perspectives. *Chem. Rev.*, 121 [10] 6124-6172 (2021)
5. F. Li, S. Zhang, D. Damjanovic, L-Q. Chen, T. R. Shrout, Local structural heterogeneity and electromechanical responses of ferroelectrics: learning from relaxor ferroelectrics. *Adv. Funct. Mater.*, 28, 1801504 (2018)
6. T. Rojac, Piezoelectric response of disordered lead-based relaxor ferroelectrics. *Comm. Mater.*, 12 (2023)
7. A. A. Bokov and Z-G. Ye, Recent progress in relaxor ferroelectrics with perovskite structure. *J. Mater. Sci.* 41, 31-52 (2006)
8. A. E. Glazounov, A. K. Tagantsev, A. J. Bell, Evidence for domain-type dynamics in the ergodic phase of the PbMg1/3Nb2/3O3 relaxor ferroelectric. *Phys. Rev. B.* 53, 11281 (1996)
9. H. Takenaka, I. Grinberg, S. Liu, A. M. Rappe, Slush-like polar structures in single-crystal relaxors. *Nature*, 546, 391-395 (2017)
10. M. Eremenko, V. Krayzman, A. Bosak, H. Y. Playford, K. W. Chapman, J. C. Woicik, B. Ravel, I. Levin, Local atomic order and hierarchical polar nanoregions in a classical relaxor ferroelectric. *Nature Comm.*, 10, 2728 (2019)
11. Kumar, J. N. Baker, P. C. Bowes, M. J. Cabral, S. Zhang, E. C. Dickey, D. L. Irving, J. M. LeBeau, Atomic-resolution electron microscopy of nanoscale local structure in lead-based relaxor ferroelectrics. *Nature Mater.*, 20, 62-67 (2021)
12. J. Hlinka, Do we need the ether of polar nanoregions. *J. Adv. Dielectrics*, 2(2) 1241006 (2012)
13. M. J. Krogstad, P. M. Gehring, S. Rosekranz, R. Osborn, F. Ye., Y. Liu, J. P. C. Ruff, W. Chen, J. M. Wozniak, H. Luo, O. Chmaissem, Z-G. Ye, D. Phelan. The relation of local order to material properties in relaxor ferroelectrics. *Nature Mater.* 17, 718-724 (2018)
14. H. Takenaka, I. Grinberg, A. M. Rappe, Anisotropic local correlations and dynamics in a relaxor ferroelectric. *Phys. Rev. Lett.* 110, 147602 (2013)
15. M. Cabral, S. Zhang, E. C. Dickey, J. M. LeBeau, Gradient chemical order in the relaxor Pb(Mg$_{1/3}$Nb$_{2/3}$)O$_3$. *Appl. Phys. Lett.*, 112 (8) 082901 (2018)
16. I. M. Reaney, J. Petzelt, V.V. Voitsekhovskii, F. Chu, N. Setter, B-site order and infrared reflectivity in A(B'B'')O$_3$ complex perovskite ceramics. J. Appl. Phys., 76, 2086-2092 (1994)
17. G. Burns and F. Dacol, Crystalline ferroelectrics with glassy polarization behavior. *Phys. Rev. B.*, 28 (5): 2527 (1983)
18. M. Otonicar, A. Bradesko, L. Fulanovic, T. Kos, H. Ursic, A. Bencan, M. J. Cabral, A. Henriques, J. L. Jones, L. Riemer, D. Damjanovic, G. Drazic, B. Malic, T. Rojac. Connecting the multiscale structure with macroscopic response of relaxor ferroelectrics. *Adv. Funct. Mater.* 30 (52) 2006823 (2020)

19. D. de Fontaine, The number of independent pair-correlation functions in multicomponent systems. *J. Appl. Cryst.* 4, 15-19 (1971)
20. In crystalline solid solutions, even a random distribution of species sharing the same crystallographic sites leads to intrinsic nanoscale clustering of unit-cell characteristics, such as composition, charge, strain, etc. In PMN-PT, a structure with a random distribution of Mg and Nb or Mg, Nb, and Ti would exhibit extended regions with net positive and negative electric charges. The chemical ordering tends to increase the size of these regions.
21. R. D. Shannon, Revised Effective Ionic Radii and Systematic Studies of Interatomic Distances in Halides and Chalcogenides. *Acta Cryst.* A32, 751 (1976)
22. A. M. Glazer, Classification of tilted octahedra in perovskites. Acta Cryst. B. 28 (11), 3384-3392 (1972)
23. I. Levin, W. J. Laws, D. Wang, I. M. Reaney, Designing pseudocubic perovskites with enhanced nanoscale polarization. *Appl. Phys. Lett.*, 111, 212902 (2017)
24. M. Matsuura, K. Hirota, P.M. Gehring, W. Chen, Z.-G. Yes, G. Shirane, Composition dependence of the diffuse scattering in the relaxor ferroelectric compound. *Phys. Rev. B* 74, 144107 (2006)
25. A. Cervellino, S. N. Gvasaliya, O. Zaharko, B. Roessli, G. M. Rotaru, R. A. Cowley, S. G. Lushnikov, T. A. Shaplygina, M. T. Fernandez-Diaz, Diffuse scattering from the lead-based relaxor ferroelectric $PbMg_{1/3}Ta_{2/3}O_3$. *J. Appl. Cryst.*, 44, 603-609
26. M. Pasciak, T. R. Welberry, J. Kulda, M. Kempa, J. Hlinka, Polar nanoregions and diffuse scattering in the relaxor ferroelectrics $PbMg_{1/3}Nb_{2/3}O_3$. *Phys. Rev. B.*, 85 (22) 224109 (2012)
27. A. Bosak, D. Chernyshov, S. Vakhrushev, M. Krisch, Diffuse scattering in relaxor ferroelectrics: true three-dimensional mapping, experimental artefacts and modelling. *J. Appl. Cryst.* 68 (1), 117-123 (2012)
28. M. Zhu, M. Xu, Y. Qi, C. Gilgenbach, J. Kim, J. Zhang, B. R. Denzer, L. W. Martin, A. M. Rappe, J. M. LeBeau, Bridging experiment and theory of relaxor ferroelectrics at the atomic scale with multislice electron ptychography. *arXiv:2408.11685*
29. A. Garcia and D. Vanderbilt, First-principles study of stability and vibrational properties of tetragonal $PbTiO_3$. *Phys. Rev.* B54 [6], 3817-3824 (1996)
30. Y. J. Wang, Y. P. Feng, Y. L. Zhu, Y. L. Tang, L. X. Yang, M. J. Zou, W. R. Geng, M. J. Han, X. W. Guo, B. Wu, X. L. Ma, Polar meron lattice in strained oxide ferroelectrics. *Nat. Mater.*, 19, 881-886 (2020)
31. L. McInnes, J. Healy, S. Astels, Hdbscan: hierarchical density based clustering. *J. Open Source Softw*. 2, 205 (2017)
32. C. Stock, L. Van Eijck, P. Fouquet, M. Maccarini, P. M. Gehring, G. Xu, J. Luo, X. Zhao, J. F. Li, D. Viehland, Interplay between static and dynamic polar correlations in relaxor $PbMg_{1/3}Nb_{2/3}O_3$. Phys. Rev. B. 81, 144127 (2010)
33. W. Dmowski, S. B. Vakhrushev, I.-K. Jeong, M. P. Hehlen, F. Trouw, T. Egami, Local lattice dynamics and the origin of the relaxor ferroelectric behavior. *Phys. Rev. Lett.*, 100, 137602 (2008)
34. A. K. Yadav, C. T. Nelson, S. L. Hsu, Z. Hong, J. D. Clakson, C. M. Schleputz, A. R. Damodaran, P. Shafer, E. Arenholz, L. R. Dedon, D. Chen, A. Vishwanath, A. M. Minor, L. Q. Chen, J. F. Scott, L. W. Martin, R. Ramesh, Observation of polar vortices in oxide superlattices, *Nature* 530, 198 (2016)

35. C. Jeong, J. Lee, H. Jo, J. Oh, H. Baik, K-J. Go, J. Son. S-Y. Choi, S. Prosandeev, L. Bellaiche, Y. Yang, Revealing the three-dimensional arrangement of polar topology in nanoparticles. *Nature Comm*., 15, 3887 (2024)
36. A. P. Levanyuk, R. Blinc. Ferroelectric phase transitions in small particles and local regions. *Phys. Rev. Lett.*, 111 (9) 097601 (2013)
37. J. Yin, H. Zong, H. Tao, X. Tao, H. Wu, Y. Zhang, L-D. Zhao, X. Ding, J. Sun, J. Zhu, J. Wu, S. J. Pennycook, Nanoscale bubble domains with polar topologies in bulk ferroelectrics. *Nature Comm.*, 12, 3632 (2021)
38. I. Luk'yanchuk, Y. Tikohonov, V. M. Vinokur, Hopfions emerge in ferroelectrics. *Nature Comm.*, 11, 2433 (2020)
39. S. L. Swartz and T. R. Shrout, Fabrication of perovskite lead magnesium niobate. *Mat. Res. Bull*. 17, 1245-1250 (1982)
40. P. Juhas, T. Davis, C. L. Farrow, S. J. L. Billinge, PDFGetX3: a rapid and highly automatable program for processing powder diffraction data into total scattering pair distribution functions. *J. Appl. Cryst*. 46, 560-566 (2013)
41. B. Ravel and M. Newwille, ATHENA, ARTEMIS, HEPASTUS: data analysis for X-ray absorption spectroscopy using IFEFFIT. *J. Synch. Rad*. 12, 537-541 (2005)
42. A. L. Ankudinov, B. Ravel, J, J, Rehr, S. D. Conradson, Real-space multiple-scattering calculations and interpretation of X-ray absorption near-edge structure. *Phys. Rev.* B58, 7565-7576 (1998)
43. A. A. Coelho, TOPAS and TOPAS-Academic: an optimization program integrating computer algebra and crystallographic objects written in C++. *J. Appl. Cryst*. 51, 201-218 (2018)
44. Certain equipment, instruments, software, or materials, commercial or noncommercial, are identified in this paper in order to specify the experimental procedure adequately. Such identification is not intended to imply recommendation or endorsement of any product or service by NIST, nor is it intended to imply that the materials or equipment identified are necessarily the best available for the purpose.
45. M. G. Tucker, D. A. Keen, M. T. Dove, A. L. Goodwin, Q. Hui, RMCProfile: reverse Monte Carlo for polycrystalline materials. *J. Phys. Cond. Matter*. 19 (33) 335218 (2007)
46. V. Krayzman, I. Levin, J. C. Woicik, Th. Proffen, T. A. Vanderah, M. G. Tucker. A combined fit of total scattering and extended X-ray absorption fine structure data for local-structure determination in crystalline materials. *J. Appl. Cryst*., 42, 867-877
47. M. Eremenko, V. Krayzman, A. Gagin, and I. Levin, Advancing reverse Monte Carlo structure refinements to the nanoscale. *J. Appl. Cryst*. 50, 1561-1570 (2017).
48. Y. Zhang, M. E. Eremenko, V. Krayzman, M. G. Tucker, and I. Levin, New capabilities for enhancement of RMCProfile: instrumental profiles with arbitrary peak shapes for structural refinements using the reverse Monte Carlo method. *J. Appl. Cryst*. 53 [6]. 1509-1519 (2020)
49. J. A. M. Paddison, Ultrafast calculation of diffuse scattering from atomisitic models. *Acta Cryst*. A. 75 [1], 14-24 (2019)
50. B. E. Warren, X-ray diffraction (Dover Publications), New York (1990).
51. R. Blinc, J. Dolinsek, A. Gregorovic, B. Zalar, C. Filipic, Z. Kutnjak, A. Levstik, and R. Pirc, Local polarization distribution and Edwards-Anderson order parameter of relaxor ferroelecrics, *Phys. Rev. Lett.*, 83, 424 (1999)

52. J. C. R. Hunt, A. A. Wray, P. Moin, Eddies, streams, and convergence zones in turbulent flows. Center for Turbulence research, 193-208 (1988)
https://web.stanford.edu/group/ctr/Summer/201306111537.pdf

**Methods**

*Sample Synthesis*

Powder specimens of (1-$x$)PMN-$x$PT with $x$=0.3, 0.35 were synthesized using the columbite route described in [39]. Powders of $MgCO_3$ and $Nb_2O_5$ (with 2 mol % excess MgO) were reacted at 900 °C for 8 h in air to form $MgNb_2O_6$, followed by another heating at 1200 C for 12 h. This product was batched with PbO (5 wt. % excess) and $TiO_2$ (<10 ppm $P_2O_5$) and heated in a covered crucible at 850 °C for 4 h and then at 1000 °C for 2h. The cover was sealed using alumina-based paste to minimize PbO loss. Prior to each heating, the powders were mixed and ground in a planetary ball mill in isopropanol using yttria-stabilized zirconia grinding media. Single crystals of these compositions were produced using a flux-growth technique, which exploits spontaneous nucleation by gradually cooling a supersaturated solution from high temperatures.

*Data Collection and Reduction*

The phase purity of the powder specimens was confirmed using a laboratory X-ray powder diffractometer equipped with Cu K$\alpha_1$ radiation. Variable-temperature neutron total scattering measurements on powder specimens were performed using POLARIS diffractometer at ISIS. The sample powders were loaded in 6 mm vanadium cans. Temperature control was achieved using a furnace. For PMN ($x$=0), we used data collected on this instrument at 300 K and 490 K previously (10). For the PMN-PT compositions, total scattering data were collected at 300 K, 383 K, and 533 K, with these temperatures selected to probe the different phase fields. Diffraction data with Bragg peaks suitable for Rietveld refinements were collected at intermediate temperatures. X-ray total scattering data were collected at the same temperatures using the I15-1 beamline at the Diamond Light Source with the incident X-ray energy of 76.6 keV and an amorphous-silicon 2D detector about 20 cm downstream from the sample. The sample powders were loaded in fused quartz capillaries, 1 mm in diameter. Temperature was controlled using cryostream and, above 500 K, with a hot-air blower. We collected data from the Si NIST SRM to calibrate the instrumental resolution function for neutron and X-ray experiments. The neutron total scattering data were reduced using the GUDRUN software to extract the neutron scattering function, S(Q), and its Fourier transform. The X-ray total scattering data were corrected for parallax effects and then reduced in the PDFGetX3 software [40] to extract the X-ray scattering function.

Variable-temperature extended X-ray absorption fine structure (EXAFS) measurements were performed for the Pb $L$3, Nb $K$, and Ti $K$ edges using the NIST 6-BM beamline at NSLS-II (Brookhaven National Laboratory). The data were collected in transmission for Pb and Nb and fluorescence for Ti. The latter was recorded using a 4-element Si-drift detector. Temperature control was achieved using a contact heater with sample powders deposited on a 10-micron-thick aluminum foil. The X-ray absorption spectra were reduced in the Athena software [41] to extract the EXAFS signal. Preliminary fitting of the EXAFS data was performed in Artemis [41], with the scattering amplitudes and phases for a photoelectron calculated using FEFF8 [42].

Single-crystal X-ray DS was measured using the beamline ID28 of the European Synchrotron Radiation Facility. Thin, rod-like crystals were mounted on a rotation stage and with a hot-air blower used for temperature control. Before the measurements, the samples were etched using hot, concentrated hydrochloric acid. DS patterns were recorded with a wavelength of 0.954 Å over the angular range of 360° in 0.1° increments on a PILATUS 1M [44] single-photon counting pixel detector. The CrysAlisPro software package was used to refine the orientation matrix, and the final reciprocal-space reconstructions of 3D diffuse-intensity distributions were accomplished using ID28-based computer software. The measured diffuse scattering was symmetrized using operations of m-3m point symmetry group to maximize the reciprocal-space coverage and mitigate possible X-ray absorption effects.

Single-crystal neutron DS data were collected on CORELLI spectrometer at the Spallation Neutron Source (SNS). These crystals were physically different from those used in the X-ray measurements but grown in the same laboratory under similar conditions. The crystals were measured at the same temperatures as used for other data. Temperature control was achieved by placing the samples in a cryo-furnace. Data processing to determine the orientation matric and obtain 3D reconstructions of the scattered intensity was performed using custom scripts developed at SNS. CORELLI is equipped with a statistical correlation chopper, which permits separating elastic scattering (energy resolution $\approx$1 meV). The 3D reconstructions were obtained with and without this chopper to compare the elastic and energy-integrated diffuse intensities. Fitting of diffuse intensity near Bragg peaks in energy-integrated 3D datasets is currently precluded by the resolution profile effects causing extended asymmetric streaks of intensity along $Q$ and towards the origin for all Bragg peaks. An additional complication is diffraction from polycrystalline aluminum foil used to shield the sample for temperature stability, which is manifested as spherical shells of intensity with the radii corresponding to the $Q$-values for the aluminum Bragg reflections.

*Structural refinements*

Rietveld refinements using the neutron powder diffraction data were performed in the TOPAS software [43, 44] to obtain average structure characteristics. This report focuses on the high-temperature cubic phases where the only refined structural variables were lattice parameters (fig. S1) and atomic displacement parameters.

A principal challenge in fitting the intensity of single-crystal DS is placing it onto the absolute scale. Bragg peaks recorded with DS are usually saturated and unsuitable for intensity calibration. The scale assigned to the diffuse intensity influences the strength of determined atomic correlations and, in the case of displacements, their magnitudes. We addressed this issue by developing a relatively rigorous and robust procedure for scaling the diffuse intensity, which is expected to yield significantly more accurate results than previously possible. In this approach, we perform spherical integration of an experimental 3D DS dataset to obtain a one-dimensional trace of intensity versus the modulus of the scattering vector $Q$. The resulting *I*($Q$) signal is compared to one obtained via a similar spherical integration of a 3D diffuse intensity distribution calculated for a sufficiently large atomic configuration. In this configuration, the atoms are displaced randomly according to atomic displacement parameters determined using Rietveld or single-crystal refinements from Bragg-peak intensities. The calculated intensities are normalized by the number of atoms in the configuration. We expect the experimental and calculated I(Q) traces to match at large Q values (Fig. S2a). Therefore, their ratio provides an estimate of the scale factor that should be applied to the experimental data. As an additional check, after rescaling and normalizing the intensity by the average scattering factor squared, we anticipate the experimental (diffuse component) and calculated traces to match the baseline in the total scattering function obtained from powders samples (Fig. S2b).

RMC refinements were performed in the RMCProfile software [45-48, 10] using simultaneous fitting of neutron and X-ray total scattering data in their reciprocal and real-space representations, neutron Bragg profile, EXAFS (Pb and Nb for PMN and Pb, Nb, and Ti for PMN-PT), and 3D reconstructions of the diffuse intensity scaled as described above (Fig. S3 through Fig. S13). The structures were represented using atomic configurations of 40 × 40 × 40 perovskite cubic unit cells, which span distances up to ≈16 nm. For the DS dataset, we used a grid with 40 × 40 × 40 voxels per reciprocal space volume bound by $H$=1, $K$=1, and $L$=1. Matching the number of voxels with the number of unit cells in the configuration was required for a Lanczos filter implemented to smooth the otherwise too-noisy calculated diffuse intensity. Unlike the previously used, more economical but aggressive, smoothing approach, the new procedure provides adequate smoothing [49] without concurrent broadening of the diffuse features, which may impact correlation lengths in the refined configuration. The DS datasets covered half of the reciprocal space with $L \geq 0$. Because of computing speed limitations, we could only include the diffuse intensity out to $H=K=\pm 2.2$, $L=2.2$ in the fits. In addition to the experimental data, we applied restraints that favored the least distorted oxygen octahedra with the target metal-oxygen distances set for each B-cation species separately according to their ionic radii and bond valence criteria. In the refined configurations, the resulting distances differed from these idealized values, being determined by the EXAFS data. Still, these restraints helped to regularize the configuration. During refinements, weights assigned to individual datasets and restraints are selected automatically following the procedure outlined in [48], which performs statistical analysis of changes in individual residual components as a function of atomic moves.

Refinement proceeded according to the following strategy. Initial configurations were generated by starting with random distributions of the octahedral metal cations (Nb and Mg for PMN and Nb, Mg, and Ti for PMN-PT), and all the atoms, including Pb and O, displaced randomly according to their ADP values (anisotropic for oxygen) obtained from our Rietveld refinements. The distributions of the B-cations in these configurations were refined in the RMCProfile software by swapping the distinct cation species among their sites while fitting the DS intensity at and around the reciprocal space $R$ points with $H$, $K$, and $L$, all half integers, which are known to reflect the short-range order of these cations. No displacement moves were allowed at this stage. Once these swap-move refinements converged, fits of the same diffuse features continued, allowing displacements of the oxygen atoms (all other atoms were left unchanged) under the polyhedral restraints to permit octahedral relaxations in response to the changes in the cation distribution. For the PMN-PT compositions, the DS contained only the first-order ½ ½ ½ superlattice reflections related to the cation ordering. These reflections were much weaker and broader compared to PMN, and they were insufficient for refining the distributions of the three B-site species. For these compositions, we combined both X-ray and neutron DS datasets, fitting the volumes of the reciprocal space that contained the superlattice reflections and their surrounding background.

For PMN, we performed independent refinements of chemical short-range order for the data collected at 300 K and 490 K and obtained similar order parameters for both cases. We then used the configuration for 490 K as a starting point for refining the 300 K structure. Refining the chemical order first instead of allowing all the atoms to move while swapping the B-cations is to avoid artifacts with the shifts of the heavy Pb and Nb species, which dominate X-ray scattering, creating an ordered pattern of displacements that reproduce the diffuse intensity. For each composition, several starting models with different distributions of the B-cations were generated following the above procedure and then used as starting models for refining atomic displacements.

A potentially limited energy integration window in the neutron total scattering measurements has been argued to be wide enough to encompass at least the low-frequency modes involving the Pb off-centering [10].

*Analyses of refined atomic configurations*

The octahedral cation distributions were characterized using the Warren-Cowley [50] short-range order parameters and the local metric introduced in [10]. This local metric, $\eta$, is defined as $\eta = (n_1 - n_2)^2 + (n_4 - n_3)^2$. Here, $n_i$ is the fraction of a given type of species (e.g., Nb) found in the $i^{th}$ coordination shell around a particular B-cation (Nb or Mg). These fractions are normalized by the expected fraction of these species based on the average composition. The metric was calculated for each B-cation in the configuration. With this definition, clusters of the B-cations with large values of $\eta$ represent chemically ordered regions.

Distortions of the oxygen octahedra, including their rotations as rigid units, were determined using a set of 18 orthonormal deformation modes described in [10]. Correlated components of atomic displacements contributing to the DS in the refined configuration were extracted via the Fourier transform of the DS complex amplitude calculated from the refined atomic coordinates (Fig. S30). The regions of reciprocal space containing the diffuse intensity of interest were selected using spherical masks, 0.2 Å$^{-1}$ in diameter. For the DS near Bragg peaks, all peaks out to $H=K=L=20$ ($Q\approx31$ Å$^{-1}$) were included – this wide coverage is computationally expensive but required to provide sufficient resolution in real space. In real space, the transform was calculated on a grid with a mesh size of 0.05 Å. The Fourier transform is real and features positive and negative peaks around atomic positions. Atoms that acquire the largest intensity of these peaks contribute the most to the DS of interest. Generally, there may be several positive peaks of varying intensity near an atomic position. A vector that connects the centers of mass for the intensity-weighted positive and negative peaks, respectively, represents the component of a displacement for a given atom that participates in the correlations, generating the diffuse intensity included in the transform. The Fourier filtering of the diffuse amplitude was accomplished using a newly developed software optimized to enable these calculations over a large volume of reciprocal space and for a large number of atoms in the configuration, as required to reveal a complete set of correlated displacements. Fig. S14 illustrates the ability of the procedure to recover correlated atomic displacements for a simulated model containing three orientational domain variants.

*Calculations of the local polarization*

The unit-cell polarization, which parallels the Edwards-Anderson local order parameter [51], depends on the cation (A – Pb or B – Mg, Ti, Nb) placed in the center of the unit cell. Therefore, we calculated the average polarization over a sampling volume of $N \times N \times N$ unit cells, sliding this volume over the configuration as in the boxcar method. The differences associated with selecting particular cation species as the center cancel already for $N=2$, facilitating the analysis. Following the trends in the average modulus of the local polarization as a function of *N* highlights the correlation length for the polarization, characterized by the number of cells required to attain the macroscopic (configuration) average. For PMN, Born effective charge (BEC) tensors have been determined using first-principles calculations [52]; however, the BEC values vary with the type of ordering for Mg and Nb. For PMN-PT, no BEC charges are available in the literature. Here, we are interested not in the absolute magnitude of the polarization but in the patterns of the local polarization vectors and the evolution of the polarization magnitude with the sampling volume. We compared the behavior of the local polarization in PMN calculated assuming the BEC values determined from first principles calculations for the [001]-type ordering (Pb – 4, Nb – 7.4, Mg – 2.6, O – -4.8 and -2.5 parallel and perpendicular to the B-O bond, respectively) and that for the polarization calculated assuming formal ionic charges for the cations. We found no significant differences in the patterns of the polarization displacement vectors for the two cases. Therefore, we adopted the formal charges to facilitate the comparison between the behavior of the local polarization in PMN and PMNT-PT solid solutions.

Fig. 2D displays trends in the modulus of the average local polarization, $\bar{P}$, as a function of the sampling volume, defined as a box of *N* × *N* × *N* unit cells. For a given *N*, the $\bar{P}$-value is obtained by averaging the moduli of the polarization within the sampling box, calculated while sliding this box over the configuration. As a reference, we included a trend for a configuration of PMN at 300 K with the refined atomic displacement vectors scrambled over the positions of each species. For this random case, the average local polarization magnitude approaches the configuration average, ⟨*P*⟩≈0 at *N* ≈ 15. For the actual configuration of PMN at 300 K, the macroscopic average is attained for *N* between 35 and 40. The decay rate for the $\bar{P}$ vs. *N* is significantly slower than in the example with random displacements, indicative of extended polar correlations. For PMN at 490 K and the PMN-PT configurations, the $\bar{P}$ falls off considerably faster but still slower than in the random case.

*Analyzing octahedral deformations and rotations*

Octahedral deformations, including rotations, were characterized using the previously developed deformation mode analysis [10]. In this method, the distortions of each octahedron are described using 18 orthonormal deformation modes, which are selected to represent typical lattice distortions and vibrational modes (see Fig. S15 in [10]). In PMN, [MgO$_6$] octahedra expand relative to the average, whereas [NbO$_6$] octahedra contract, as expected from the differences in the ionic radii of Mg and Nb. A strong positive correlation (Pearson correlation coefficient 0.2) exists between the local chemical ordering parameter and the magnitude of this "breathing mode," which is consistent with the regions of stronger chemical ordering permitting larger oxygen shifts along the Mg-O-Nb bonds as required to satisfy the bonding requirements of both Mg and Nb without an overall strain. Likewise, in the PMN-PT configurations, the [MgO$_6$] octahedra expand, whereas both [NbO$_6$] and [TiO$_6$] octahedra contract with the much more significant shrinkage of the latter consistent with the Ti ionic radius being smaller than that of Nb.

In PMN, analysis of octahedral tilting modes around the cubic axes reveals a cogwheel pattern of rotations ordered over a short range within octahedral layers normal to the corresponding rotation axis (Fig. 1D). The average rotation angle about each cubic axis is about 4°, and the correlation is limited to the first two coordination shells. At 490 K, the correlations are significant only for the nearest-neighbor octahedra. Analysis of correlations along the octahedral chains parallel to the rotation axes highlights a weak preference for the in-phase tilts. That is, the PMN structure at 300 K appears to contain small clusters, 1-2 nm in size, with $a^0a^0c^+$-type tilting. At 490 K, the cluster size is reduced to just the neighboring octahedra. The magnitude of octahedral rotations correlated positively with the local chemical order parameter for the cations; however, the chemical and in-plane tilting order parameters appeared uncorrelated. No ordering of rotations was detected for the PMN-PT compositions.

*Vector-field metrics*

*Q*-criterion

Hunt *et al*. [52[ have proposed this metric as effective in distinguishing regions of coherent rotation motion from those dominated by strain. While originally developed for fluid flows, it is applicable to displacement vector fields for identifying vortex-like or rotational structures.

We denote a displacement vector field as $\boldsymbol{u}(x,y,z) = \big(u_x(x,y,z), u_y(x,y,z), u_z(x,y,z)\big)$, where $\boldsymbol{u_x}, \boldsymbol{u_y}, \boldsymbol{u_z}$ are the components of the vector field along the x, y, and z axes, respectively.

The spatial gradient of $\boldsymbol{u}$, $\nabla \boldsymbol{u}$, is a second-order tensor:

$$\nabla \boldsymbol{u} = \begin{pmatrix} \frac{\partial u_x}{\partial x} & \frac{\partial u_x}{\partial y} & \frac{\partial u_x}{\partial z} \\ \frac{\partial u_y}{\partial x} & \frac{\partial u_y}{\partial y} & \frac{\partial u_y}{\partial z} \\ \frac{\partial u_z}{\partial x} & \frac{\partial u_z}{\partial y} & \frac{\partial u_z}{\partial z} \end{pmatrix}$$

This tensor captures all first-order spatial derivatives of the vector field, combining information about both deformation and rotation. The gradient tensor can be decomposed into a symmetric strain-rate tensor **S** and a skew-symmetric rotation tensor **Ω**, defined as

$$\boldsymbol{S} = \frac{1}{2}(\nabla \boldsymbol{u} + (\nabla \boldsymbol{u})^T)$$

$$\boldsymbol{\Omega} = \frac{1}{2}(\nabla \boldsymbol{u} - (\nabla \boldsymbol{u})^T),$$

where

$$\|\boldsymbol{S}\|^2 = \sum_{i,j} S_{ij}^2$$

$$\|\boldsymbol{\Omega}\|^2 = \sum_{i,j} \Omega_{ij}^2.$$

Then, the Q-criterion is defined as a local measure of excess rotation relative to strain:

$$Q = \frac{1}{2}(\|\boldsymbol{\Omega}\|^2 - \|\boldsymbol{S}\|^2)$$

If Q>0, rotation dominates over strain, indicating a vortex or vortex-like region. Alternatively, if Q<0, strain dominates.

Curl:

The curl is another metric describing the tendency of a vector field to exhibit the local rotation. It is a vector having a direction aligned with the axis of rotation and a magnitude reflecting the rotational strength. The curl is defined as:

$$\nabla \times \mathbf{u} = \left( \frac{\partial u_z}{\partial y} - \frac{\partial u_y}{\partial z}, \frac{\partial u_x}{\partial z} - \frac{\partial u_z}{\partial x}, \frac{\partial u_y}{\partial x} - \frac{\partial u_x}{\partial y} \right)$$

*Divergence*

The divergence measures the flow of the vector field from a given point. It is a scalar field defined as:

$$\nabla \cdot \mathbf{u} = \frac{\partial u_x}{\partial x} + \frac{\partial u_y}{\partial y} + \frac{\partial u_z}{\partial z}$$

Physically, a positive divergence indicates an expansion region, while a negative divergence indicates a compression region.